\documentclass[aps,twocolumn]{revtex4-1}
\usepackage{amsmath}
\usepackage{hyperref}
\usepackage{fullpage}
\usepackage{graphicx}

\renewcommand{\vec}[1]{\mathbf{#1}}
\begin{document}

\title{Electrical conductivity and resonant states of doped graphene
considering next-nearest neighbor interaction}
\author{J.E. Barrios-Vargas and Gerardo G Naumis}
%{\it Depto. de F\'{i}sica-Qu\'{i}mica, Instituto de F\'{i}sica, Universidad Nacional Aut\'onoma de M\'exico (UNAM).
%Apdo. Postal 20-364, 01000, M\'exico D.F., M\'exico.}
\email{naumis@fisica.unam.mx}
\affiliation{Depto. de F\'{i}sica-Qu\'{i}mica, Instituto de F\'{i}sica, Universidad Nacional Aut\'onoma de M\'exico (UNAM).
Apdo. Postal 20-364, 01000, M\'exico D.F., M\'exico.}
\begin{abstract}
The next-nearest neighbor interaction (NNN) is included in a tight-binding
calculation of the electronic spectrum and conductivity of doped graphene.
As a result, we observe a wide variation of the conductivity behavior, since
the Fermi energy and the resonance peak are not shifted by the same amount.
Such effect can have a profound effect in the idea of explaining the minimal
conductivity of graphene as a consequence of impurities or defects. Finally,
we also estimate the mean free path and relaxation time due to resonant
impurity scattering.\\

\noindent PACS number(s): 81.05.ue, 72.80.Vp, 73.22.Pr
\end{abstract}

\maketitle

%%%%%%%%%%%%%%%%%%%%%%%%%%%%%%%%%%%%%%%%%%%%%%%%%%%%%%%%%%%%%%%%%
% Abstract
%%%%%%%%%%%%%%%%%%%%%%%%%%%%%%%%%%%%%%%%%%%%%%%%%%%%%%%%%%%%%%%%%

%%%%%%%%%%%%%%%%%%%%%%%%%%%%%%%%%%%%%%%%%%%%%%%%%%%%%%%%%%%%%%%%%
%Introduction
%%%%%%%%%%%%%%%%%%%%%%%%%%%%%%%%%%%%%%%%%%%%%%%%%%%%%%%%%%%%%%%%%

\section{Introduction}

Research on graphene has seen a rush of publications since its experimental
discovery in 2004 \cite{novoselov}. The vast fame of this carbon allotrope 
\cite{geimgra} is partly due to its amazing room-temperature transport
properties \cite{peres}, as for example the high electronic \cite{novoselov2}
and thermal conductivity \cite{balandin}, profiling nano-devices based on
graphene \cite{avouris}. From the theoretical viewpoint, graphene is also
astonishing because the charge carriers are described by massless Dirac
fermions \cite{novoselov2,katsnelson} as a consequence of the crystal
symmetry. However, in the construction of electronic nano-devices, the use of
pure graphene present some troubles. For example, the conductivity is
difficult to manipulate by means of an external gate voltage, which is a
desirable feature required to build a FET transistor. Such performance is
related to the Klein paradox in relativistic quantum mechanics \cite%
{katsnelson}, or from a more standard outlook, as a consequence of the zero
band gap. There are many proposals to solve the problem. For instance, by
using quantum dots \cite{silvestrov}, a graphene nanomesh \cite{bai}, an
external electromagnetic radiation source \cite{lopez,lopez2} or by doping
using impurities \cite{naumis}. In fact, in a previous article we showed
that impurities lead to a metal-insulator transition since a mobility edge
appears near the Fermi energy \cite{naumis}. Such prediction has been
confirmed in graphene doped with H \cite{bostwick}, which opens the
possibility to build graphene-based narrow gap semiconductors \cite{naumis}.
Other groups have shown that graphene can present an n-typed semiconductor
behavior when doped with N, Bi or Sb atoms; and a p-type semiconductor
behavior using B or Au atoms \cite{wei,gierz}. Still, there is much debate
regarding the nature of the mobility transition, since in 2 dimensional (2D)
scaling theory, it is predicted that all states are localized in the
presence of a finite amount of disorder \cite{amini,schleede}.\\

The appearance of a mobility edge has its origins in the presence of
resonant states when few impurities are considered \cite{naumis}. Both type
of states, localized and resonant, have an enhanced amplitude in the
neighborhood of the impurity. Nevertheless, resonant states only trap
electrons during a short time. Using a nearest-neighbor (NN) tight-binding
model, resonant states have been reported near the Fermi energy \cite{pereira}. 
Furthermore, an analytical approximate expression was found for
the resonant energy as a function of the impurity energy using the Lifshitz
equation \cite{skrypnyk,skrypnyk2}. The main difference between such works
is that Pereira {\it et al.} also noted that there is a slightly
different between the resonance energy obtained from the Lifshitz equation,
and the actual localization of the sharp resonance in the density of states (%
$DOS$) when the impurity energy is not so strong \cite{pereira}. It is
necessary to remark that only strong impurities are able to produce resonant
states \cite{wehling}. The presence of next-nearest neighbor interaction
(NNN) shifts the Fermi energy and breaks the electron-hole symmetry \cite{castrorev}. 
Including NNN interaction, the $DOS$ displays a sharp peak when
a vacancy is considered like an impurity in the lattice \cite{pereira2}.
Moreover, this peak is smeared by the NNN.\\

The central topic of this work is to emphasize the different behaviors in
the electrical conductivity due to resonances when NNN interactions are
included, as happens in carbon nanotubes \cite{zhangcnt}. Usually, the NNN
interaction is not taken into account in graphene tight-binding calculations
\cite{roche1}, so here we propose a systematic study of the subject. Such study is
important because there is a debate concerning which mechanisms determine
the charge carrier mobility \cite{wehling2,roche2}, as well as the nature of
the minimal conductivity \cite{peres}. As we will see, an impurity can
produce a sharp peak or a smoothening effect in the electrical conductivity,
depending on the charge doping, temperature, strength of the impurity
scattering and the value of the NNN interaction. The interplay between such
factors is subtle since for example, the Fermi level and the resonance
energy are not shifted by the same amount when NNN interaction is included.
It is worthwhile mentioning that the electrical conductivity at high
temperatures is determined basically by the electron-phonon interaction \cite%
{yuan}, while here we discuss only scattering by impurities. Thus, our
results are relevant only for low temperatures. However, this case is
important to explain the weak temperature dependence of the conductivity,
which is basically proportional to the carrier concentration \cite%
{novoselov2} \cite{ni}.\newline

The layout of the work is the following. In section \ref{secmodel} we
describe the model and the perturbative approach used to calculate Green's
function for a NNN tight binding Hamiltonian of doped graphene. Section \ref%
{secgreen} describes the calculation of Green's function of pure graphene,
which is used in Section \ref{secimpurity} to calculate the resonant
energies. Section \ref{conduc} contains the electrical conductivity
calculations using the Kubo-Greenwood formula. Finally, in section \ref%
{seccon} we present the conclusions. 
%%%%%%%%%%%%%%%%%%%%%%%%%%%%%%%%%%%%%%%%%%%%%%%%%%%%%%%%%%%%%%%%%
%Model
%%%%%%%%%%%%%%%%%%%%%%%%%%%%%%%%%%%%%%%%%%%%%%%%%%%%%%%%%%%%%%%%%

\section{Model}

\label{secmodel}

As a model, we consider a pure graphene tight-binding Hamiltonian with
substitutional impurities at very low concentrations. Since there are no
correlations between impurities and the impurity concentration is very low,
we can reduce the problem to a single localized impurity in a graphene
lattice. The behavior for a given low concentration can be found by a simple
implementation of the virtual crystal approximation (VCA) \cite{economou}.
Also, we will use the fact that the graphene's honeycomb lattice is formed
by two triangular interpenetrating sublattices, denoted A and B \cite%
{castrorev}. The corresponding tight-binding Hamiltonian is,

\begin{align*}
\mathcal{H}=\mathcal{H}_{0}+\mathcal{H}_{1} \,;
\end{align*}
\begin{align}  \label{eqHam}
& \mathcal{H}_{0}=-t\sum_{\langle i|j \rangle,\sigma} \left(
a^{\dagger}_{\sigma,i}b^{\phantom{x}}_{\sigma,j} +b^{\dagger}_{\sigma ,j}a^{%
\phantom{x}}_{\sigma,i} \right)  \notag \\
& -t^{\prime}\sum_{\langle\langle i|j \rangle\rangle,\sigma} \left(
a^{\dagger}_{\sigma,i}a^{\phantom{x}}_{\sigma,j} +b^{\dagger}_{\sigma ,i}b^{%
\phantom{x}}_{\sigma,j} +a^{\dagger}_{\sigma,j}a^{\phantom{x}}_{\sigma,i}
+b^{\dagger}_{\sigma,j}b^{\phantom{x}}_{\sigma,i} \right) \,,
\end{align}
\begin{align*}
\mathcal{H}_{1}=\varepsilon\left( a^{\dagger}_{\sigma,l}a^{\phantom{x}%
}_{\sigma,l} \right) \quad\mathrm{{or} \quad\mathcal{H}_{1}=\varepsilon%
\left( b^{\dagger}_{\sigma,l}b^{\phantom{x}}_{\sigma,l} \right) \,.}
\end{align*}
where $a^{\phantom{x}}_{\sigma,i}$ ($a^{\dagger}_{\sigma,i}$) annihilates
(creates) an electron with spin $\sigma$ ($\sigma=\uparrow,\downarrow$) on
site $i$ at position $\vec{R}_{i}$ on the A sublattice (an equivalent
definition is used for B sublattice), $t\,(\approx2.79\mathrm{eV})$ is the
NN hopping energy, and $t^{\prime}\,(\approx0.68\mathrm{eV})$ is the NNN
hopping energy \cite{rubio}. $\varepsilon$ is the energy difference between
a carbon atom and a foreign atom, and $l$ is the impurity position. \newline

Usually, the resonances are characterized by looking at the Green's
functions ($G$) of $\mathcal{H}$. Expressing $G$ as a perturbation series in
terms of $G_{0}$ (which is the Green's function corresponding to the
unperturbed Hamiltonian $\mathcal{H}_{0}$), a closed expression is obtained
for the local density of states ($LDOS$) in the impurity site $l$ \cite%
{economou}, 
\begin{align}  \label{eqLDOS}
\rho(l;E)=\frac{\rho_{0}(l;E)}{|1-\varepsilon G_{0}(l,l;E)|^{2}} \,,
\end{align}
where $G_{0}(l,l;E)$ and $\rho_{0}(l;E)$ are respectively the Green's \
function and $LDOS$ on $l$ site with $\mathcal{H}_{0}$.\newline

The term $|1-\varepsilon G_{0}(l,l;E)|^{2}$ cannot become zero for $E$
within the band. However, under certain values of $\varepsilon$, this term
is near to zero for a given $E\approx E_{\mathrm{r}}$. Then, a sharp peak in
the $LDOS$ will emerge around $E_{\mathrm{r}}$. This $E_{\mathrm{r}}$ is
associated with a resonant state inasmuch as there is a different impurity
energy level. If $\mathrm{Im}\left\{ G_{0}(l,l;E)\right\} $ is a slowly
varying function of $E$ (for $E$ around $E_{\mathrm{r}}$), then the resonant
energy will be given as a solution of the Lifshitz equation, 
\begin{align}  \label{eqEr}
1-\varepsilon\mathrm{Re}\left\{ G_{0}(l,l;E)\right\} \approx0\,.
\end{align}
Furthermore, if the derivative of $\mathrm{Re}\left\{ G_{0}(l,l;E)\right\} $
does not have a strong dependence of $E$ near $E_{\mathrm{r}}$, then \cite%
{economou}, 
\begin{align*}
\frac{1}{|1-\varepsilon G_{0}(l,l;E)|^{2}}\sim\frac{\Gamma^{2}}{(E-E_{%
\mathrm{r}})^{2}+\Gamma^{2}}\,,
\end{align*}

where $\Gamma$ corresponds to the width of the impurity resonance, 
\begin{align}  \label{eqGamma}
\Gamma=\frac{|\mathrm{Im}\left\{ G_{0}(l,l;E_{\mathrm{r}})\right\} |}{|%
\mathrm{Re}\left\{ G_{0}^{^{\prime}}(l,l;E)\right\} |}\,.
\end{align}

Thus, the resonant state effect is sketched by its location, $E_{\mathrm{r}}$
in equation~\eqref{eqEr}; and its width, $\Gamma$ in equation~\eqref{eqGamma}. 
Those characteristics of the resonant state are inherited from the
Green's function behavior, which is presented in the section below.

%%%%%%%%%%%%%%%%%%%%%%%%%%%%%%%%%%%%%%%%%%%%%%%%%%%%%%%%%%%%%%%%%
%Green's function
%%%%%%%%%%%%%%%%%%%%%%%%%%%%%%%%%%%%%%%%%%%%%%%%%%%%%%%%%%%%%%%%%

\section{Green's function of pure Graphene with NNN interaction}

\label{secgreen}

To solve the Lifshitz equation \eqref{eqEr}, we need to obtain the Green's
function for graphene with NNN interaction. Notice that analytical
expressions are available only for the NN interaction \cite{horiguchi}, and
not for the NNN interaction. The Green's function can be obtained from, 
\begin{align}  \label{eqGreen}
G_{0}(E) = \left[ \frac{1}{N}\sum_{\vec{k}\in\mathrm{1BZ}} \frac{1}{E+is-E(%
\vec{k})}\right] \,,
\end{align}
where $s\ll1$ and $E(\vec{k})$ is the dispersion relationship of equation~%
\eqref{eqHam}, and given by \cite{castrorev}, 
\begin{align}  \label{eqEk}
E_{\pm}(\vec{k})=\pm\sqrt{3+f(\vec{k})}-t^{\prime}f(\vec{k})\,,
\end{align}
\begin{align*}
f(\vec{k})=2\cos\left( \sqrt{3}k_{y}a\right) +4\cos\left( \frac{\sqrt{3}}{2}
k_{y}a\right) \cos\left( \frac{3}{2}k_{x}a\right) \,,
\end{align*}
where the minus sign applies to the valence and the plus sign the conduction
band. Around the Dirac point ($\vec{K}$ or $\vec{K}^{\prime}$), the momentum
can be written as $\vec{k}=\vec{K}+\vec{q}$, where $\vec{q}$ is a small
vector. Then, equation~\eqref{eqEk} up to second order is given by \cite%
{castrorev}, 
\begin{align}
E_{\pm}(\vec{q})\approx & 3t^{\prime}\pm\frac{3ta}{2}|\vec{q}|  \notag \\
& -\left\{ \frac{9t^{\prime}a^{2}}{4}\pm\frac{3ta^{2}}{8}\sin\left[ 3\left(
\arctan\frac{q_{x}}{q_{y}}\right) \right] \right\} |\vec{q}|^{2}\,,
\label{eqEq}
\end{align}
where $a$ is the carbon-carbon distance ($a\approx1.42$\AA ).\newline

\begin{figure}[h]
\centering
\begin{minipage}{0.48\textwidth}
\begin{tabular}{l}
\includegraphics[width=1.0\textwidth]{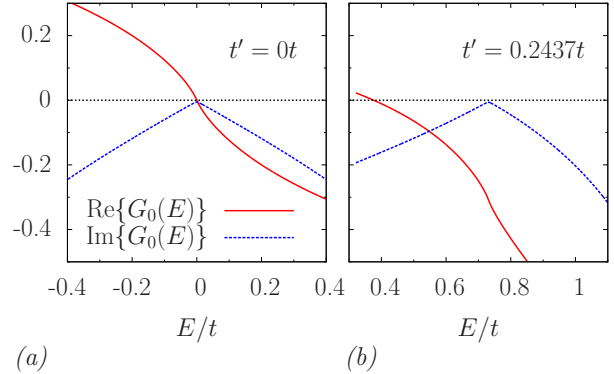}
\end{tabular}
\end{minipage}
\caption{(\textit{a}) Green's functions with nearest neighbor interaction
and (\textit{b}) with next-nearest neighbor interaction. [The square mesh in
the first Brillouin zone that was used to calculate $G_{0}(l,l;E)$ using the
equation~\eqref{eqGreen} is uniform and contains $N=7.5\times10^{7}$ points,
and $s=2\times10^{-3}$.]}
\label{fig:Green}
\end{figure}

To compute equation~\eqref{eqGreen} we used a square mesh in the first
Brillouin zone of the reciprocal space to evaluate the sum. The results are
presented in figure~\ref{fig:Green}. In figure~\ref{fig:Green}\textit{(a)}
we show the result when the NNN behavior is absent ($t^{\prime}=0$). The
result is in excellent agreement with the analytical formula \cite{horiguchi}%
. Notice how at zero energy (corresponding to the Fermi energy for pure
graphene, $E_{\mathrm{F}}^{0}$), the imaginary and real part of the Green's
function cross at zero energy, resulting in a symmetrical behavior for $%
\varepsilon$ around $E=0$, equation~\eqref{eqEr}. This symmetry is broken
when $t^{\prime }\neq0$, as seen in figure~\ref{fig:Green}\textit{(b)}. This
is due to the fact that real part of the Green's function no longer cross
the zero at $E_{\mathrm{F}}^{0}=3t^{\prime}$, this energy value matches with
the zero value of the imaginary part.

%%%%%%%%%%%%%%%%%%%%%%%%%%%%%%%%%%%%%%%%%%%%%%%%%%%%%%%%%%%%%%%%%
%Single impurity
%%%%%%%%%%%%%%%%%%%%%%%%%%%%%%%%%%%%%%%%%%%%%%%%%%%%%%%%%%%%%%%%%

\section{Green's function of a single impurity in graphene}

\label{secimpurity}

Once the Green's function $G_{0}$ for pure graphene is known, we can compute 
$G$ in order to describe the resonant states. In the following subsections
we present the corresponding results. It is useful keeping in mind that the
impurity effect is mainly weighted by the real part of $G_{0}$.

\subsection{Local Density of States ($LDOS$)}

\begin{figure}[h]
\begin{minipage}{0.48\textwidth}
\begin{tabular}{ll}
\includegraphics[width=1.0\textwidth]{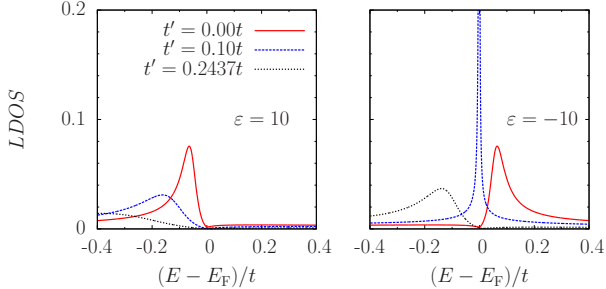}

\end{tabular}
\end{minipage}
\caption{$LDOS$ calculated using equation~\eqref{eqLDOS} for different
values of $\protect\varepsilon$. and $t^{\prime}$. [The parameters used to
generate the graphs are the same as those used in Figure~\protect\ref%
{fig:Green}.]}
\label{fig:LDOS}
\end{figure}

The $LDOS$ can be calculated using equation~\eqref{eqLDOS}, since 
\begin{align*}
\rho_{0}(l;E)=-\frac{1}{\pi}\mathrm{Im}\left\{ G_{0}(l,l;E)\right\} \,.
\end{align*}
Using the calculated $G_{0}$, and considering strong impurities, i.e. $%
\varepsilon/t>3$, in figure~\ref{fig:LDOS} we can see that the $LDOS$
exhibits a peak at certain resonant energies for two combinations of
impurities self-energies $\varepsilon$ and different NNN interaction $%
t^{\prime}$. An evident characteristic is that the sharp location, i.e. the
resonant energy, has a shift depending on the $t^{\prime}$ parameter, this
is a consequence of the shift in the ordinate axis of $\mathrm{Re} \left\{
G_{0}(l,l;E) \right\} $, as observed in figure~\ref{fig:Green}. Another
characteristic that corresponds to the sharpest $LDOS$ behavior emerges when
the $E_{\mathrm{r}}$ is near the $E_{\mathrm{F}}^{0}$. From figure~\ref%
{fig:LDOS}, it is clear that the NNN interaction radically changes the
resonance properties when compared with the NN case. Therefore, the NN
interaction is not enough to describe the behavior of the doped system. To
see this in more detail, let us calculate the position of the resonant
energy and the resonance width.

%%%%%%%%%%%%%%%%%%%%%%%%%%%%%%%%%%%%%%%%%%%%%%%%%%%%%%%%%%%%%%%%%

\subsection{Resonant energy ($E_{\mathrm{r}}$)}

\begin{figure}[tbp]
\begin{minipage}{0.4\textwidth}
\begin{tabular}{c}
\includegraphics[width=\textwidth]{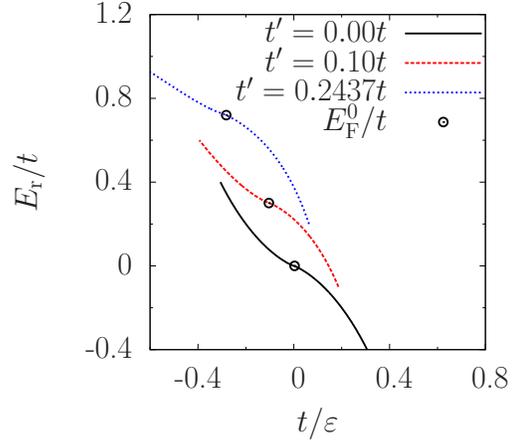}
\end{tabular}
\end{minipage}
\caption{The $E_{\mathrm{r}}=E_{\mathrm{r}}(\protect\varepsilon )$ curve
defined by the Lifshitz equation~\eqref{eqEr}, for different values of NNN
interaction $t^{\prime }$. The Fermi energy for pure graphene including NNN
interaction, $E_{\mathrm{F}}^{0}$, is identified by a circle. [The
parameters used to generate the graphs are the same as those used in Figure~%
\protect\ref{fig:Green}.]}
\label{fig:Ereps}
\end{figure}

In order to obtain $E_{\mathrm{r}}$, we need to solve the Lifshitz equation~%
\eqref{eqEr}. As a result, in the figure~\ref{fig:Ereps} we present the
energies $E_{\mathrm{r}}$ that satisfy equation~\eqref{eqEr} as a function
of $\varepsilon$ for different sets of $t^{\prime}$. The main effect of the
NNN interaction, is a shift proportional to $t^{\prime}$, as expected from
the first correction to the NN interaction in equation~\eqref{eqEq}.
However, also the curvature of $E_{\mathrm{r}}(\varepsilon)$ in figure~\ref%
{fig:Ereps} exhibits a slight difference as the NNN hopping energy varies.
Additionally, in the same figure, we plot the Fermi energy $E_{\mathrm{F}%
}^{0}$ for pure graphene including NNN interaction for certain $\varepsilon$
at which the Fermi energy lies exactly at the resonance $LDOS$ peak. In
other words, the circles in figure~\ref{fig:Ereps} are the values of the
parameters $\varepsilon$ and $t^{\prime}$ where the electronic properties
are most affected, since electrons at the Fermi level have the exact energy
of a resonant state and can thus be easily trapped for a certain amount of
time around the impurity.

According with Skrypnyk \cite{skrypnyk}, for the NN interaction, the
resonant states that are near $E_{\mathrm{F}}^{0}$, in the asymptotic limit $%
\varepsilon \rightarrow \pm \infty $ are given by, 
\begin{equation*}
\frac{1}{\varepsilon }\propto E_{\mathrm{r}}\ln \left\vert E_{\mathrm{r}%
}\right\vert \,. 
\end{equation*}%
In such limit, we can extend the previous expression to include the NNN
interaction as follows, 
\begin{equation*}
\frac{t}{\varepsilon }\approx \frac{5}{2\sqrt{3}\pi }\left( \frac{E_{\mathrm{%
r}}-3t^{\prime }}{t}\right) \ln \left\vert \frac{E_{\mathrm{r}}-3t^{\prime }%
}{t}\right\vert \,. 
\end{equation*}%
(Notice that the numerical factor in the above expression was not reported
by Skrypnykv \textit{et al.} \cite{skrypnyk} since their expression used for
the dispersion relation was not normalized). \newline
Although the previous expressions follows form a rigid translation of the
spectrum with $t^{\prime }$, the more realistic cases are those of small $%
\varepsilon $, in which the effects of $t^{\prime }$ are important, since
the resonance peak is not shifted by the same amount. As we will see, this
effect is important when one considers the effects on the electronic
conductivity.

%%%%%%%%%%%%%%%%%%%%%%%%%%%%%%%%%%%%%%%%%%%%%%%%%%%%%%%%%%%%%%%%%

\subsection{Resonant width}

\begin{figure}[ptb]
\begin{minipage}{0.4\textwidth}
\begin{tabular}{c}
\includegraphics[width=\textwidth]{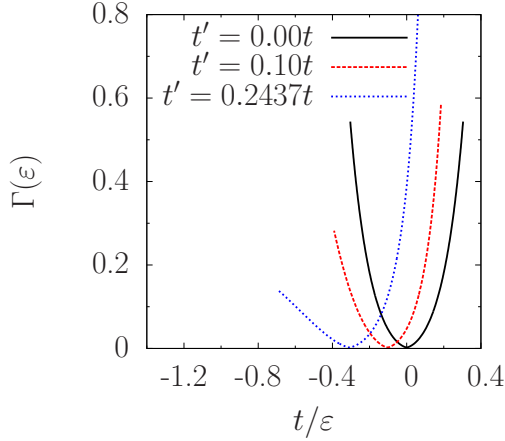}
\end{tabular}
\end{minipage}
\caption{Resonant width, $\Gamma$, given by equation~\eqref{eqGamma} as a
function of the impurity effect, $\protect\varepsilon^{-1}$. The figure
shows asymmetric curves due to the NNN interaction. [The parameters used to
generate the graphs are the same as those used in Figure~\protect\ref%
{fig:Green}.]}
\label{fig:Gamma}
\end{figure}

In figure~\ref{fig:Gamma} we present the influence of the NNN interaction $%
t^{\prime}$ on the resonance width $\Gamma$ as a function of $\varepsilon
^{-1}$. It is evident the influence of the NNN due to the asymmetry in the
curves for $t^{\prime}\neq0$. Observe that the asymmetry leads to a reduced
resonance width for $\varepsilon<0$ when the NNN interaction increases. This
means that the peak is shaper and thus, the lifetime of the resonance is
increased.\newline

On the contrary, the opposite is observed for $\varepsilon>0$; \textit{i.e.}%
, the lifetime of the electron near the impurity is decreased by the NNN
interaction.\newline

As was previously mentioned, the most notorious peak in the $LDOS$ is
located near the Fermi energy. The corresponding $\varepsilon$ value to that
energy is clearly observed in figure~\ref{fig:Gamma}, which is the graph
corresponding to equation~\eqref{eqGamma}.

%%%%%%%%%%%%%%%%%%%%%%%%%%%%%%%%%%%%%%%%%%%%%%%%%%%%%%%%%%%%%%%%%
%DC conductivity
%%%%%%%%%%%%%%%%%%%%%%%%%%%%%%%%%%%%%%%%%%%%%%%%%%%%%%%%%%%%%%%%%

\section{DC conductivity}

\label{conduc} Resonances are important to the transport properties, as for
example, in the electrical or thermal conductivity. In this section, we
evaluate the electrical conductivity ($\sigma _{xx}$) taking into account
resonant states and the NNN interaction. To do so, we use the Kubo-Greenwood
formula expressed as \cite{economou}, 
\begin{equation}\label{KG}
\sigma _{xx}=\frac{e^{2}\hbar }{\pi m^{2}}\frac{N}{\Omega _{0}}\int_{-\infty
}^{\infty }dE\,\,\,\mathcal{T}(E)\left( \frac{\partial f}{\partial E}\right) %
\bigg|_{E+\mu }\,, 
\end{equation}%
where, 
\begin{equation*}
\mathcal{T}(E)=\mathrm{Tr}\left\{ p_{x}\mathrm{Im}\left\{ G(E)\right\} p_{x}%
\mathrm{Im}\left\{ G(E)\right\} \right\} \,, 
\end{equation*}%
and $\Omega _{0}=3a^{2}$ is the primitive cell area, $f$ is the Fermi-Dirac
distribution and $\mu $ is the chemical potential, which can be tuned by the
external field (for example, with a voltage applied in the lattice). $p_{x}$
is the momentum operator, given by the following commutator, 
\begin{equation*}
p_{x}=\frac{im}{\hbar }[\mathcal{H},x]\,. 
\end{equation*}%
It is necessary to remark that here, the Hamiltonian operator includes the
next-nearest neighbor interaction. Therefore, $p_{x}$ inherits this
interaction. Aftermath, $p_{x}$ can be written in terms of the momentum
operator associated to the NN interaction as follows. Consider first the
momentum $p_{x}$ for NN, 
\begin{align}\label{px}
p_{x}^{\mbox{{\tiny NN}}}& =\frac{imt}{\hbar }[x,\mathcal{W}]\,, \nonumber \\
& =\frac{imt}{\hbar }\sum_{l=1}^{N}\sum_{m\in \mbox{\tiny NN}}(\vec{R}_{l}-%
\vec{R}_{m})_{x}\mathcal{W},
\end{align}%
where we introduced the connectivity matrix defined as, 
\begin{equation*}
\mathcal{W}(m,n)=\left\{ 
\begin{array}{cc}
1 & \mbox{if }m\mbox{ and }n\mbox{ are NN} \\ 
0 & \mbox{otherwise}%
\end{array}%
\right. \,. 
\end{equation*}%
Using this connectivity matrix, the Hamiltonian without perturbation
including the NNN interaction can be rewritten as, 
\begin{equation*}
\mathcal{H}_{0}=-t\mathcal{W}-t^{\prime }(\mathcal{W}^{2}-3\mathcal{I})\,, 
\end{equation*}%
where $\mathcal{I}$ is the identity matrix. Taking the previous expression
and using \eqref{px}, we obtain the corresponding operator for the NNN case, 
\begin{align}
p_{x}^{\mbox{{\tiny NNN}}}& =\frac{im}{\hbar }[-t\mathcal{W}-t^{\prime }(%
\mathcal{W}^{2}-3\mathcal{I}),x]\,,  \notag \\
& =p_{x}^{\mbox{{\tiny NN}}}+\frac{t^{\prime }}{t}\bigg(p_{x}^{%
\mbox{{\tiny
NN}}}\mathcal{W}+\mathcal{W}p_{x}^{\mbox{{\tiny NN}}}\bigg)\,.
\end{align}

\begin{figure}[h]
\begin{minipage}{0.48\textwidth}
\begin{tabular}{ll}
\includegraphics[width=1.0\textwidth]{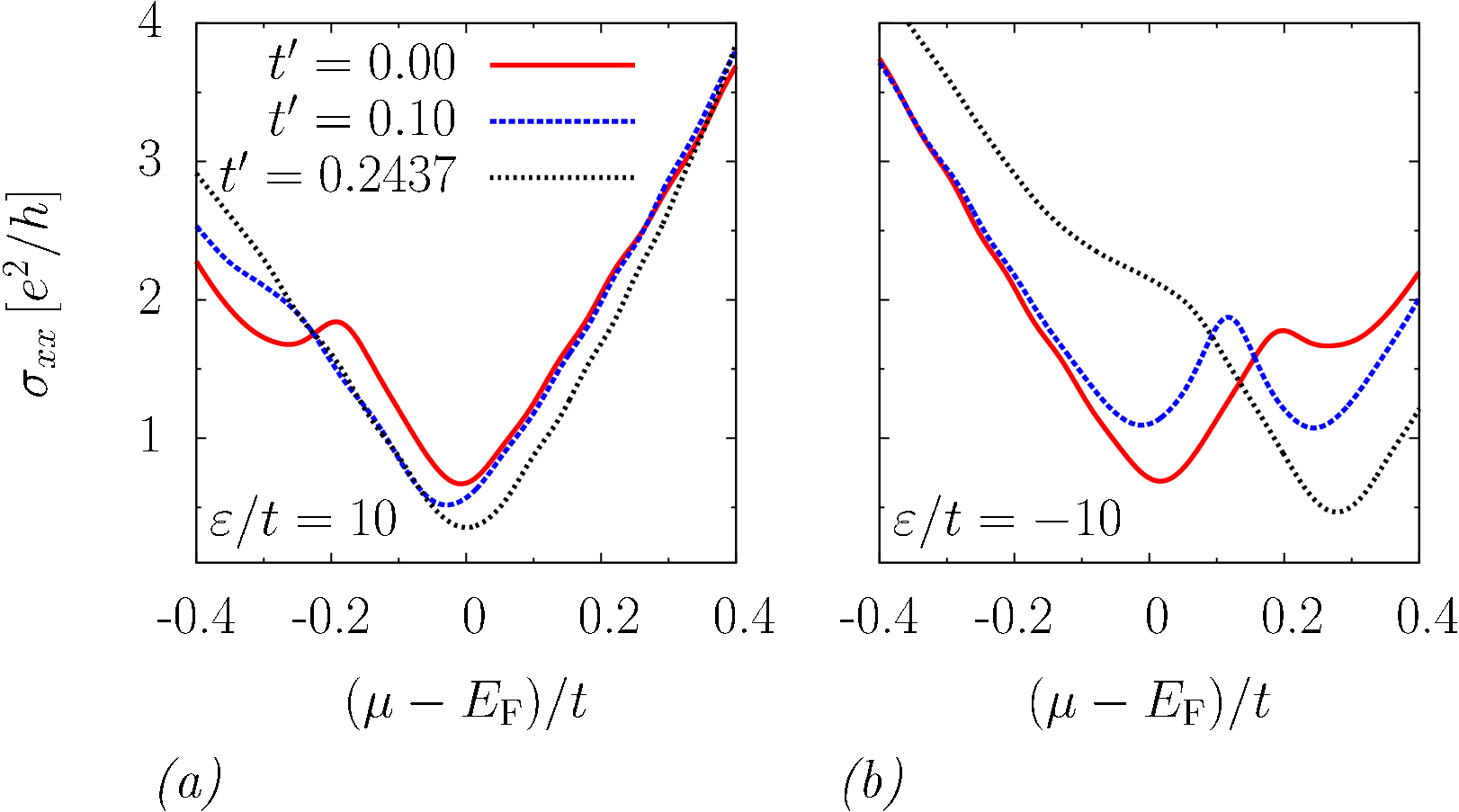}
%&\includegraphics[width=0.5\textwidth]{imagenes/figure6b.pdf}\\
%({\it a})&({\it b})
\end{tabular}
\end{minipage}
\caption{Electrical conductivity calculated using the Kubo-Greenwood formula
at low concentration of impurities, $C=0.01$. \textit{(a)} Corresponds to $%
\protect\varepsilon/t=10$ and \textit{(b)} corresponds to $\protect%
\varepsilon/t=-10$. [All lattices have $N\approx10^{4}$.]}
\label{fig:sigma}
\end{figure}

The previous expressions are effortlessly written in a computer program, in
which we consider a low concentration of impurities, $C$, introduced adding
the perturbation $\mathcal{H}_{1}$ with the same $\varepsilon $ at different
sites taken at random with a uniform distribution. \ In figure~\ref%
{fig:sigma} we show the conductivity calculated using the Kubo-Greenwood
formula \eqref{KG}. Considering different values of $t^{\prime }$ and as a
function of the charge doping. Figure~\ref{fig:sigma} was made at a fixed
representative temperature, in this case $k_{\mathrm{B}}T=0.025\mbox{ eV}$%
, to highlight the main effects of the NNN interaction. Clearly, figure~\ref%
{fig:sigma} exhibits the radical difference in the conductivity behavior due
to the NNN interaction, since an smearing effect, as seen in figure~\ref%
{fig:sigma}\textit{(a)}, can appear, or as in figure~\ref{fig:sigma}\textit{%
(b), }a sharp peak can be observed . In fact, if we look at the temperature
behavior, we can also get very different behaviors of $\sigma _{xx}$. These
changes are due to a subtle interplay between the chemical doping,
temperature, the NNN interaction and the impurity type.\newline

\begin{figure}[h]
\begin{minipage}{0.48\textwidth}
\begin{tabular}{c}
\includegraphics[width=\textwidth]{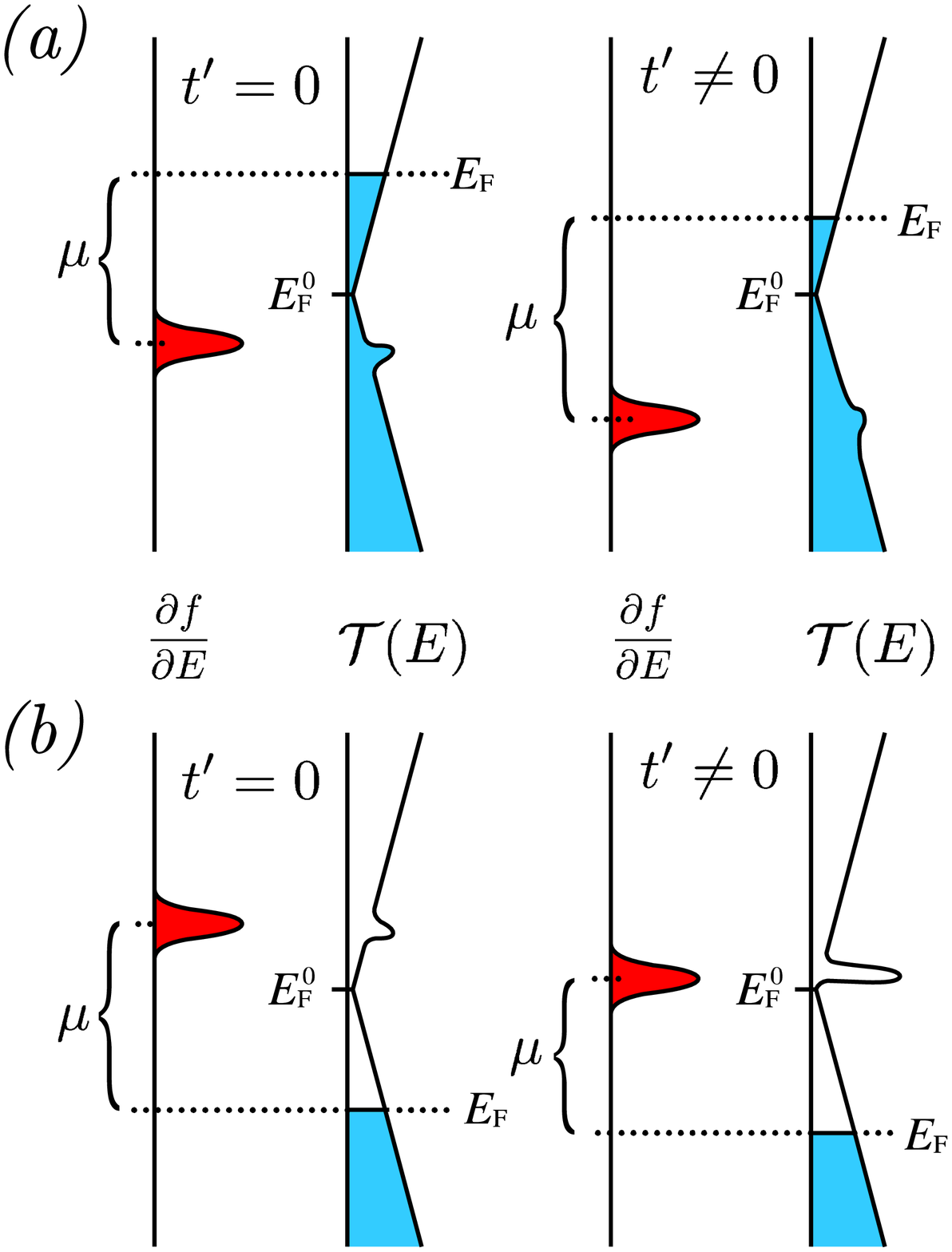}
\end{tabular}
\end{minipage}
\caption{Schematic diagram of the Kubo-Greenwood formula which explains the
effect of the NNN interaction. The behavior of the two building blocks, the
thermal selector of states $\partial f/\partial E$ and the trace $\mathcal{T}%
(E)$, are shown on each panel. The position of the resonance energy and
Fermi energy are also shown. The following cases are considered: \textit{(a)}
$\protect\varepsilon/t=10$ and \textit{(b)} $\protect\varepsilon/t=-10$.
Both graphics assume the case $t^{\prime}/t=0.1$. }
\label{fig:muvar}
\end{figure}

To understand the diverse behaviors of $\sigma_{xx}$, in figure~\ref%
{fig:muvar} we show a sketch of the "building blocks" that appear in the
Kubo-Greenwood, and how such blocks are modified by the considered
parameters. On each panel of the graph, at the left we show the shape of the
term $\partial f/\partial E$, which corresponds to the "thermal selector".
At $T=0$, it becomes a delta function centered at the chemical potential ($%
\mu$), while at $T\neq0$ has a width of the order of $k_{\mathrm{B}}T$. The
position of $\partial f/\partial E$ on the energy axis can be externally
modified by doping with charge carriers, resulting in different positions of
the Fermi energy $(E_{\mathrm{F}}-\mu)$ when compared with the equilibrium
value of such energy, denoted by $E_{\mathrm{F}}$. The second building block
is $\mathcal{T}(E)$, which can be thought as a quantum selector of the
transport channels. Since the Green's function of doped graphene can be
written as,%
\begin{align}  \label{eqGperturbed}
G\approx G_{0}+\sum_{l}\frac{G_0|l\rangle\varepsilon \langle l| G_0}{%
1-\varepsilon G_0(l,l)}
\end{align}
(where the sum is carried over impurity sites), $\mathcal{T}(E)$ has two
behaviors. Near the resonant energies, $G$ is dominated by the second term
in Eq. (\ref{eqGperturbed}), and $G_{0}$ can be \ neglected. As a
consequence, a peak appears in the conductivity at the resonant energy, as
shown in figure~\ref{fig:muvar}. Far from the resonance, $G\approx G_{0}$.
Then we recover the transmittance of pure graphene. \newline

Now we can study how the two building blocks interact to produce many
different behaviors. First, it is clear that variations in $\mu $ and $T$
can produce a peak in the conductivity if the thermal selector coincides
with the resonance peak. So for example, the conductivity can be enhanced if
for example, at a certain temperature the thermal selector begins to have an
overlap over the \ peak of $\mathcal{T}(E)$. The effect of the NNN
interaction is very subtle since in principle one can expect a simple
translation in energy of the spectrum. However, as it was said in the
previous sections, the rigid translation of the spectrum is only valid at
high $\varepsilon$. According to our results, for
realistic impurities there are deviations from such behavior, and thus, $E_{%
\mathrm{F}}$ and $E_{\mathrm{r}}$ are not rigidly translated, i.e., the
distance $\left\vert E_{\mathrm{F}}-E_{\mathrm{r}}\right\vert $ depends on
the NNN interaction. Since the conductivity depends a lot on such
factor due to the position of the thermal selector, the resulting effect of
the NNN happens to be very important. One can see that in fact, the most
important factor is the position of the resonant peak, which can be changed
due to different kinds of impurities. Thus, one can expect a wide
variability of the conductivity due to resonant scattering, as is observed
in nanotubes, wheres even similar samples can present wide variations of the
conductivity \cite{zhangcnt}. \newline

Finally, if we consider the scattering term in Eq. \eqref{eqGperturbed}, the
impurities allow to define a relaxation time, $\tau_s$, as follows. 
\begin{align*}
\frac{1}{\tau_s}=\frac{N_{\mathrm{imp}}}{N} S\,,
\end{align*}
where $S$ is the scattering probability per unit time. $S$ can be calculated
by summing over all transition rates from $|i\rangle\to |f\rangle$ , 
\begin{align*}
s_{if}=f_i(1-f_f)W_{if} \,,
\end{align*}
using the Fermi golden rule, 
\begin{align*}
W_{if}=\frac{2\pi}{\hbar} | \langle f| Q |i\rangle |^2 \delta (E_f-E_i) \,,
\end{align*}
where 
\begin{align*}
| \langle f| Q |i\rangle |^2&=\frac{\varepsilon^2}{|1-\varepsilon
G_0(l,l;E)|^2}  \\
&\approx \frac{1}{\pi^2 \rho_0^2(l;E_{\mathrm{r}})} \frac{\Gamma^2}{(E-E_{%
\mathrm{r}})^2+\Gamma^2} \,.
\end{align*}

After steps similar to those used to get the Kubo-Greenwood formula \cite{economou},
we obtain, 
\begin{equation}\label{eqS}
S=\frac{2\pi k_{\mathrm{B}}T}{\hbar }\int_{-\infty }^{\infty }dE\,\,\,%
\mathcal{Q}(E)\left( \frac{\partial f}{\partial E}\right) \,, 
\end{equation}%
where, 
\begin{equation*}
\mathcal{Q}(E)\approx \frac{\rho ^{2}(E)}{\rho _{0}^{2}(l;E_{\mathrm{r}})}%
\frac{\Gamma ^{2}}{(E-E_{\mathrm{r}})^{2}+\Gamma ^{2}}\,. 
\end{equation*}%
Far from the resonance peak, $E_{\mathrm{r}}$, and near $E_{\mathrm{F}}^{0}$, 
\begin{equation*}
\rho (E)\approx \frac{1}{\sqrt{3}\pi }\frac{|E-E_{\mathrm{F}}^{0}|}{t^{2}}\,, 
\end{equation*}%
then the scattering term $\mathcal{Q}(E)$ is given by, 
\begin{equation}\label{eqQfarEr}
\mathcal{Q}(E)\approx \frac{1}{3\pi ^{2}}\frac{1}{\rho _{0}^{2}(l;E_{\mathrm{%
r}})}\frac{\Gamma ^{2}}{(3t'-E_{\mathrm{r}})^{2}}\frac{|E-3t'|^{2}}{t^{4}}\,. 
\end{equation}\\

At $T=0$, the resulting relaxation time obtained from a straightforward
evaluation of Eq. \eqref{eqS} using \eqref{eqQfarEr} and remembering that $E_{\rm F}^0=3t'$,
we obtain
\begin{equation*}
\tau _{s}^{-1}\approx \frac{4C}{3h}\frac{k_{\mathrm{B}}T}{\rho _{0}^{2}(l;E_{%
\mathrm{r}})}\frac{\Gamma ^{2}}{(3t'-E_{\mathrm{r}})^{2}}%
\frac{|E_{\mathrm{F}}-3t'|^{2}}{t^{4}}\,. 
\end{equation*}\\

Thus, the mean free path ($l$) is $l\approx v_{\mathrm{F}}\tau _{s}$, which
goes as $E_{\mathrm{F}}^{-2}$.\\

%%%%%%%%%%%%%%%%%%%%%%%%%%%%%%%%%%%%%%%%%%%%%%%%%%%%% %%%%%%%%%%%%
%Conclusions
%%%%%%%%%%%%%%%%%%%%%%%%%%%%%%%%%%%%%%%%%%%%%%%%%%%%%%%%%%%%%%%%%

\section{Conclusions}\label{seccon}

We have studied the effects in the spectrum and electronic conductivity of
low concentrations of impurities in graphene when the next-nearest neighbor
interaction is considered in a tight-binding approximation. Although the
electronic spectrum is basically similar to the case of pure nearest
neighbor interaction, the conductivity is \ much more affected since the
Fermi level and the resonance peak are not shifted by the same amount,
resulting in a wide variability of the conductivity, as happens with carbon
nanotubes \cite{zhangcnt}. As a consequence, the present study shows that is
difficult to explain the minimum electrical conductivity of graphene at the
Dirac point by only considering impurity scattering, and thus the evanescent
waves approach seems to be a possible explanation \cite{peres}. Also, we
obtained that the relaxation time in graphene goes like $E_{\mathrm{F}}^{-2}$, 
resulting in a large mean free path.\newline

%%%%%%%%%%%%%%%%%%%%%%%%%%%%%%%%%%%%%%%%%%%%%%%%%%%%%%%%%%%%%%%%%
%Acknowledgments
%%%%%%%%%%%%%%%%%%%%%%%%%%%%%%%%%%%%%%%%%%%%%%%%%%%%%%%%%%%%%%%%%
\acknowledgments
%\ack
We thank the DGAPA-UNAM project IN-1003310-3. Calculations were performed on
Kanbalam and Bakliz supercomputers at DGSCA-UNAM.

%\pagebreak
%%%%%%%%%%%%%%%%%%%%%%%%%%%%%%%%%%%%%%%%%%%%%%%%%%%%%%%%%%%%%%%%%
%Bibliography
%%%%%%%%%%%%%%%%%%%%%%%%%%%%%%%%%%%%%%%%%%%%%%%%%%%%%%%%%%%%%%%%%
\bibliographystyle{apsrev4-1}
\bibliography{biblioImpEff}

\end{document}